\documentclass[aps,prb,twocolumn, superscriptaddress]{revtex4}

\usepackage{graphicx}

\begin{document}


\title{ Anisotropic phonon conduction and lattice distortions

in CMR-type bilayer manganite
(La$_{1-z}$Pr$_{z}$)$_{1.2}$Sr$_{1.8}$Mn$_{2}$O$_{7}$ 

($z$=0,0.2,0.4 and 0.6) single crystals}


\author{ M.Matsukawa} 
\email{matsukawa@iwate-u.ac.jp }
\author{ M.Narita}
\author{ T.Nishimura}
\author{ M.Yoshizawa }
\affiliation{ Department of Materials Science and Technology, Iwate University Morioka 020-8551 , Japan }
\author{ M. Apostu}
\author{ R. Suryanarayanan }
\author{ A. Revcolevschi }
\affiliation{ Laboratoire de Physico-Chimie de L'etat Solide, CNRS, UMR 8648
 Universite Paris-Sud, 91405 Orsay, France }
\author{ K. Itoh }
\affiliation{ National Institute for Materials Science, Tsukuba 305-0003, Japan }
\author{ N. Kobayashi }
\affiliation{ Institute for Materials Research, Tohoku University, Sendai  980-8577, Japan }

\date{\today}

\begin{abstract}
We have undertaken a systematic study of thermal conductivity as a function of temperature and magnetic field of single crystals of the compound (La$_{1-z}$Pr$_{z}$)$_{1.2}$Sr$_{1.8}$Mn$_{2}$O$_{7}$ for $z$(Pr) =0.2,0.4. and 0.6. The lattice distortion due to Pr-substitution and anisotropic thermal conductivity  in bilayer manganites are discussed on the basis of different relaxation models of  local lattice distortions in metal and insulating states proposed by Maderda et al.  The giant magnetothermal effect  is scaled as a function of magnetization and discussed  on the basis of a systematic variation of the occupation of the $e_g$-electron orbital states due to Pr-substitution. 
\end{abstract}


\maketitle

\section{INTRODUCTION}
Since the discovery of colossal magnetoresistance (CMR) in the hole-doped perovskite manganites Ln$_{1-x}$D$_{x}$ MnO$_{3}$ (Ln = trivalent rare earth ; D = divalent alkaline earth Ca,Sr,Ba,Pb)  , extensive studies of the field-induced complex physics exhibited by these CMR compounds have been reported. \cite{TO00} Though the double exchange(DE) mechanism \cite{ZE51} plays a major role in explaining some of the properties, a quantitative model to account for the insulator-metal transition and the CMR effect is yet to be proposed. In the DE model, the carriers are strongly ferromagnetically coupled, due to Hund's Rule coupling, to the Mn core which also results in the hopping of electrons between the Mn ions. However, it was pointed out that the coupling of carriers to local Jahn-Teller(JT) distortions and to the Mn spins must also be considered\cite{MI95}. Further, several recent experiments seem to support the phase separation model\cite{DA01}. Thermal transport measurements provide additional crucial information on the various scattering mechanisms of thermal carriers such as electrons and phonons and are expected to improve the understanding of some of the phenomena encountered in the manganites. There have been relatively few reports \cite{VI97,CO97,UH98,HE99} on thermal transport in cubic manganites. The thermal conductivity of cubic manganites has been  discussed on the basis of  static and dynamic lattice distortions by several authors. It should be noted that the Ln$_{1-x}$D$_{x}$ MnO$_{3}$ cubic perovskite compounds belong to the Ruddlesden-Popper series generally described as (Ln,D)$_{n+1 }$Mn $_{n}$O$_{3n+1}$  with $n =\infty$ . For $n=2$, one obtains the bilayer Mn perovskite (Ln,D) $_{3}$ Mn$_{2}$O$_{7}$,in which two MnO$_{2}$ layers are stacked with (Ln,D)$_{2}$O$_{2}$  layers along the $c$-axis of the structure. The reduced dimensionality of these compounds has been shown to have interesting consequences on the physical properties. Let us note that the optimally hole doped compound La$_{1.2}$Sr$_{1.2}$Mn$_{2}$O$_{7}$ ($x$=0.4) exhibits a CMR of more than 98\% near the paramagnetic insulator (PI) to ferromagnetic metal (FM) transition temperature $T_{c}$ =$\sim$120K.\cite{MO96}  The thermal conductivity measurements of the double layer manganite, La$_{1.2}$Sr$_{1.2}$Mn$_{2}$O$_{7}$ were reported recently indicating  an enhanced  giant magnetothermal  effect  due to its layered structures \cite{MA00}.
For the end member $x=0$ LaMnO$_{3}$ ,which is a cubic compound with  JT distortions, it was reported in that  the value of $\kappa$  measured on a stoichiometric single crystal  corresponds to a thermally good conductor , like an ideal phonon gas \cite{ZH01}.  On the other hand,  the end member $x=1$ compound , CaMnO$_{3}$, without JT active Mn sites, shows a higher  thermal conduction  in comparison with  the low $\kappa$  data obtained for the hole doped compounds \cite{HE99}.  Therefore, it is desirable to examine a close relationship existing between  the lattice distortion and thermal conductivity  in layered manganites. 

In this paper, we report a systematic study of thermal conductivity as a function of temperature and magnetic field of single crystals of the compound (La$_{1-z}$Pr$_{z}$)$_{1.2}$Sr$_{1.8}$Mn$_{2}$O$_{7}$, for $z$(Pr) =0.2,0.4 and 0.6.    Keeping the hole doping fixed at $x=0.4$  , the chemical pressure effect due to substitution of Pr on La-ion sites  enhances anisotropic lattice parameters \cite{OG00,AP01} and results in a variation of the $e_{g}$-electron character, as reported by Moritomo et al \cite{MO97}.   The lattice distortion due to Pr-substitution and the anisotropic thermal conductivity  in  bilayer manganites were discussed on the basis of different relaxation models of  local lattice distortions in metal and insulating states proposed by Medarde et al.\cite{ME99}.  Next,  the giant magnetothermal resistivity   was scaled as a function of magnetization and discussed  on the basis of a systematic variation in $e_{g}$-electron orbital states. 
\section{EXPERIMENT}
Single crystals of 
(La$_{1-z}$Pr$_{z}$) $_{1.2}$Sr$_{1.8}$Mn$_{2}$O$_{7}$  ($z$=0,0.2,0.4 and 0.6) were grown from sintered rods of the same nominal composition by the floating-zone method, using a mirror furnace. Crystals could be easily cleaved to yield shiny surfaces. X-ray Laue patterns have indicated the cleaved surface to be along the $ab$-plane of the structure and therefore the $c$-axis perpendicular to it. An X-ray powder pattern taken on a small part of the cleaved crystal did not indicate the presence of any additional phases. The calculated lattice parameters are listed in previous works \cite{OG00,AP01}. The Pr-substitution at La ion sites resulted in a small contraction of the $a$-axis parameter and an expansion of the $c$-axis parameter.  Energy-dispersive X-ray analyses of the cleaved surface revealed the compositions of the two samples to be close to those of the respective sintered rods. The dimensions of the crystals used in the experiments reported here were typically 3.4$\times$ 2.8mm$^2$ in the $ab$-plane and 1.0 mm along the $c$-axis.  

The end member compound Sr$_{3}$Mn$_{2}$O$_{7}$ was prepared  by a powder sintering technique under the annealing condition  1650 $^\circ$C for 12hours first reported by Mizutani et al., \cite{MI70} and by Mitchell et al., \cite{MI99}.  After rapidly cooling it down to room temperatures to prevent decomposition,  the polycrystalline sample was post-annealed at 400 $^\circ$C  for 24hours to remove its oxygen-deficiency. The dimensions of the cylindrical sintered sample used in the thermal measurement were 4.4mm in diameter and 8mm along the longitudinal direction.  The X-ray powder pattern of this sample  showed that the sample obtained consists mainly of a single-phase with the 327 crystal structure. The magnetization measurement also revealed that the present sample shows an anti-ferromagnetic transition around $T_{N}$=$\sim$160K \cite{MI99}. 
The thermal conductivity ,both in the $ab$-plane and along the $c$-axis of the crystal, was measured by means of the steady-state heat-flow method using a G-M type cryocooler in zero field. Magnetothermal conductivity measurements were carried out at the National Research Institute for Metals and at the High Field Laboratory for Superconducting Materials, Institute for Materials Research, Tohoku University.  The resistivity in the $ab$-plane of the crystal was measured using a conventional four-probe technique. The zero-field-cooled (ZFC) magnetization measurements were made using a SQUID magnetometer both at Iwate University and in Orsay. 
\section{RESULTS AND DISCUSSION}
\begin{figure}
\includegraphics[width=10cm]{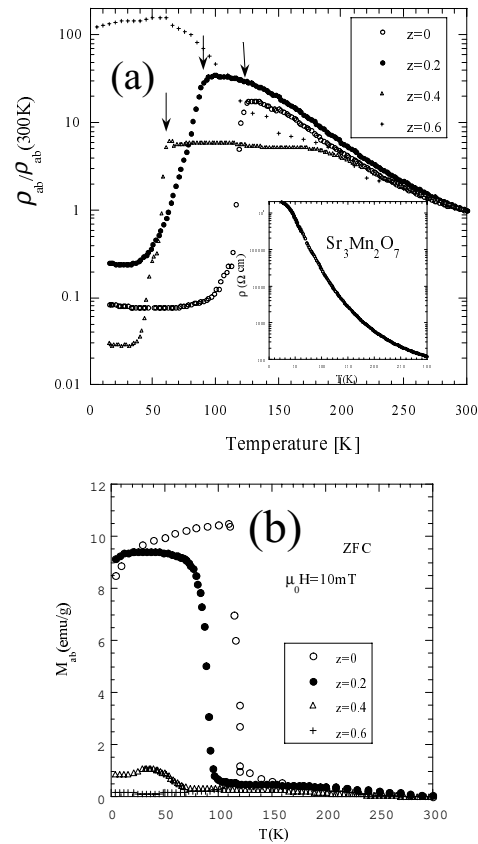}%
\caption{\label{RMT} (a)$ab$-plane normalized resistivity $\rho$$_{ab}$ and (b) $ab$-plane magnetization $M_{ab}$ for (La$_{1-z}$Pr$_{z}$) $_{1.2}$Sr$_{1.8}$Mn$_{2}$O$_{7}$ ($z=0$, 0.2, 0.4 and 0.6) single crystals. The arrows denote the PI to FM transition temperature. The $\rho$ data  of the polycrystalline sample Sr$_{3}$Mn$_{2}$O$_{7}$ are also shown in the inset of Fig.1 (a).}
\end{figure}%

Figures \ref{RMT} (a) and (b) present the $ab$-plane normalized resistivity $\rho$$_{ab}$ and the $ab$-plane magnetization $M_{ab}$ of our samples ($z=0$, 0.2, 0.4 and 0.6)  as a function of temperature.  The $\rho$ data  of the polycrystalline sample Sr$_{3}$Mn$_{2}$O$_{7}$ are also shown in the inset of Fig.1 (a).   For the $z=0$ crystal, the value of $\rho$$_{ab}$ dropped sharply at $\sim$120 K, close to the insulating to metallic transition ,with decreasing temperature. This temperature was reduced to $\sim$60K in the sample with $z=0.4$.  Finally, the $z=0.6$ sample showed neither an insulator to metal transition nor a paramagnetic to ferromagnetic transition.   The temperature variation of  the  resistivity  of the $x=1$ sample showed a negative curvature over the whole measured range and  the value of  $\rho$ rapidly increased by  more than four orders of magnitude, up to $2\times10^{6}$ $\Omega$cm at 20K  from $\sim$100 $\Omega$cm at room temperature (RT) ,  which is  in sharp contrast with the typical value of  $\sim$0.1 $\Omega$cm at RT for Pr-substituted samples \cite{OG00,AP01}.   Recently , a theoretical band calculation \cite{ME02} predicts  that the bilayer perovskite Sr$_{3}$Mn$_{2}$O$_{7}$ is  an antiferromagnetic insulator  with an indirect gap of 0.45 eV  , which  is qualitatively in good agreement with our magnetization and resistivity data. Moreover,   the electronic structure of  this compound is expected to be similar to   that of  the cubic perovskite end member CaMnO$_{3}$. 

\begin{figure}
\includegraphics[width=10cm]{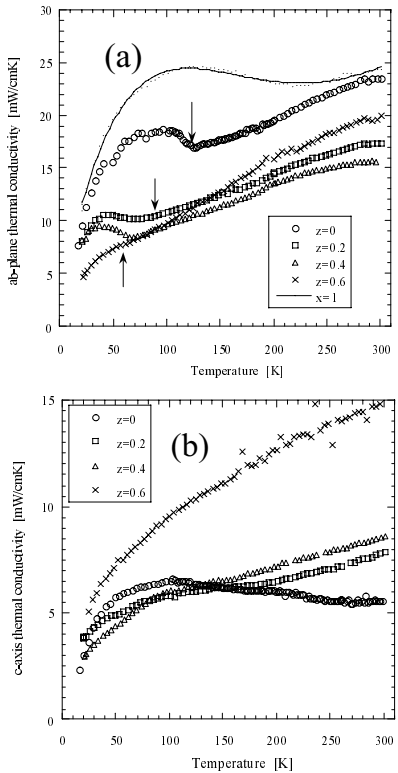}%
\caption{\label{KT} (a)$ab$-plane and (b)$c$-axis thermal conductivity  of  (La$_{1-z}$Pr$_{z}$) $_{1.2}$Sr$_{1.8}$Mn$_{2}$O$_{7}$  ($z$=0, 0.2 ,0.4 and 0.6) single crystals. The arrows denote the PI to FM transition temperature. For comparison, the temperature variation of  polycrystalline Sr$_{3}$Mn$_{2}$O$_{7}$  is also presented in Fig.2(a).}
\end{figure}%

The $ab$-plane and the $c$-axis thermal conductivity  of  (La$_{1-z}$Pr$_{z}$) $_{1.2}$Sr$_{1.8}$Mn$_{2}$O$_{7}$  ($z$=0, 0.2 ,0.4 and 0.6) single crystals are  shown in Fig.\ref{KT} as a function of temperature, up to 300K. The arrows denote the PI to FM transition temperature. For comparison, the temperature variation of the end member polycrystalline sample Sr$_{3}$Mn$_{2}$O$_{7}$  is also presented in Fig.2(a).  We note, first that a low thermal conductivity was observed over a wide of temperature range for all samples ,in spite of the high-quality of our single-crystals. Second, the temperature dependence of $\kappa$ in the $ab$-plane exhibited anomalies associated with the PI to FM transition, while no such anomalies were observed for $\kappa$ along the $c$-axis direction near $T_{C}$ for any of the samples.  Above $T_{C}$, the value of $\kappa_{ab}$ showed a positive curvature but tended towards a constant value near room temperatures with increasing temperatures. 
The electron thermal conductivity ,for the  $z=0$ sample ,is estimated  to be at most $\sim$1 \% of the total value, even in the metallic  state from the resistivity data based on the Wiedemann-Franz (W-F) law  and reproduce, neither the whole anomaly in $\kappa_{ab}$, nor the giant thermal effect around $T_{C}$ \cite{MA00}.  Thus, heat carriers are taken as phonons for all measured samples. Here,  the phonon mean free path $l_{ph}$ in the insulating state of the $x=0.4$ sample is estimated from the specific heat data $C_{ph}$, and  longitudinal sound  velocity $v_{ph}$ , using the kinetic expression for the phonon gas  $\kappa=1/3\cdot C_{ph}v_{ph}l_{ph}$.
If   $C_{ph}$(150K)=1.78 J/cm$^{3}$K \cite{VA01} and $\kappa_{ab}$(150K)=11.3 mW/cm (present data) for the 
(La$_{1-z}$Pr$_{z}$) $_{1.2}$Sr$_{1.8}$Mn$_{2}$O$_{7}$  single crystal ($z=0.4$)  and $v_{ph}$=4.3$\times10^{5}$ cm/s \cite{BR00} in the $ab$-plane  for the La $_{1.2}$Sr$_{1.8}$Mn$_{2}$O$_{7}$  bilayer single crystal ,  we then obtain  $l_{ph,ab}$ =4.43 \AA \ at 150K which is comparable with the Mn-O-Mn bond length ($\sim$4 \AA).  In a smilar way, the value of $l_{ph,ab}$ reaches 4.07 \AA \ at 300K.  In our estimation, the value of $v_{ab}$ is taken as a constant  since the relative variation of the longitudinal velocity is very small from ultrasonic  measurements on the bilayer manganites. 
 
Next, the end member compound Sr$_{3}$Mn$_{2}$O$_{7}$   showed a different behavior from the $ab$-plane data  in  the compound (La$_{1-z}$Pr$_{z}$) $_{1.2}$Sr$_{1.8}$Mn$_{2}$O$_{7}$.  The value of $\kappa$ of the $x=1$ sample remained almost a constant down to $\sim$ 200K, then showed a slight increase upon crossing $T_{N}=\sim$ 160 K and  finally  dropped rapidly with decreasing temperature.  The temperature dependence of $\kappa$ of the $x=1$ sample reminds us of a typical phonon conduction in insulating materials, except for the magnitude of measured values. The extremely high resistivity of the $x=1 (n=2)$ sample ( $\sim$100 $\Omega$cm at RT) in sharp contrast with the value of $\rho_{ab}$ ($\sim0.1\Omega$cm at RT) for the Pr-substituted $x=0.4 (n=2)$  samples gives rise to a negligible contribution of small polarons ,which probably results in a typical phonon conduction.  In comparison with the thermal data of the polycrystalline sample of Sr$_{3}$Mn$_{2}$O$_{7}$ , low thermal values of single crystal samples are anomalous indicating a strong phonon damping which correlates with dynamical lattice distortions coupled with thermally hopping of localized carriers at Mn-sites (as discussed later). 
\begin{figure}
\includegraphics[width=10cm]{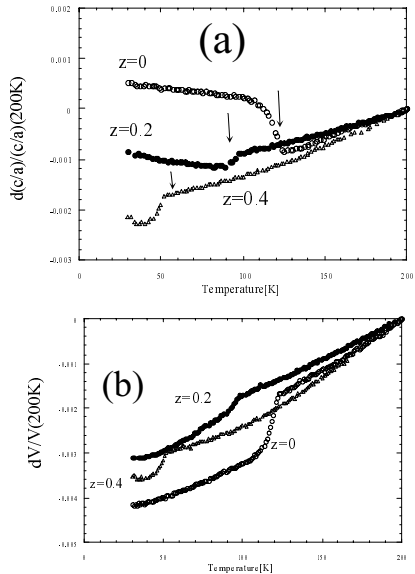}%
\caption{\label{LT}(a)anisotropic lattice striction and (b)volume striction,as a function of temperature, up to 300K, for (La$_{1-z}$Pr$_{z}$) $_{1.2}$Sr$_{1.8}$Mn$_{2}$O$_{7}$ ($z=0$, 0.2, 0.4 and 0.6) single crystals. The arrows denote the PI to FM transition temperature.}
\end{figure}%

We discuss the relationship between  local lattice distortions and thermal conductivity  in bilayer manganites.   Figures \ref{LT} (a) and (b) show the anisotropic lattice striction and volume striction for the z=0, 0.2 and 0.4 samples, respectively.  The anomaly in anisotropic lattice striction reverses sign near z=0.2 but  the volume of all samples rapidly shrinks upon crossing $T_{C}$. These findings have been discussed in ref.21 
on the basis of different relaxation models of  local lattice distortions in metal and insulating states proposed by Maderda et al.\cite{ME99}. The authors tried to explain the different behaviors of average and local lattice distortions observed in neutron powder diffraction experiments on  
La$_{2-2x}$Sr$_{1+2x}$Mn$_{2}$O$_{7}$ (0.32$\leq$x$\leq$0.4).   Based on their explanation, the observed volume shrinkage near $T_{C}$ seems to arise from  a strong suppression  in the local lattice disorder  manifested by Mn$^{3+}$-O$_{6}$ and  Mn$^{4+}$-O$_{6}$ octahedra with the JT distorted and undistorted sites  when the system is going towards the metallic state from the insulating state.  The MnO bond-length distribution becomes narrower below $T_{C}$ because of a screening effect  due to itinerant carriers.  Accordingly , the phonon conduction is limited by the local lattice disorder accompanying  a polaron hopping in the insulating state.  However, in the metallic region , it is expected that  the homogenous structure of the local lattice associated with carrier delocalization, causes an upturn in the $ab$-plane thermal conductivity, below $T_{C}$ .  It should be noted that this conclusion is  reached  from the lattice striction and thermal data of the same samples, combined with the preceding model.  Furthermore,  in comparison with  the anisotropic Debye-Waller (D-W) factor of Sr$_{3}$Mn$_{2}$O$_{7}$  ,  the anisotropic D-W data of 
La$_{2-2x}$Sr$_{1+2x}$Mn$_{2}$O$_{7}$ (0.32$\leq$x$\leq$0.4)  show that, even in the metallic  state of the latter, the inhomogeneities in the local lattice structure still remain.  This finding is probably related, not only with a small upturn in $\kappa$, but also to the low values of $\kappa$. 
On the other hand,  the out-of-plane phonon conduction across the MnO double-layers is strongly disturbed from stacking fault-like scattering , of the layered structures.  Phonon scattering due to stacking faults dominates thermal conduction along the $c$-axis,   masks some contribution of  the local lattice distortion associated with the insulator to metal transition and results in to the anisotropy  of the  thermal conductivity.  
\begin{figure}
\includegraphics[width=10cm]{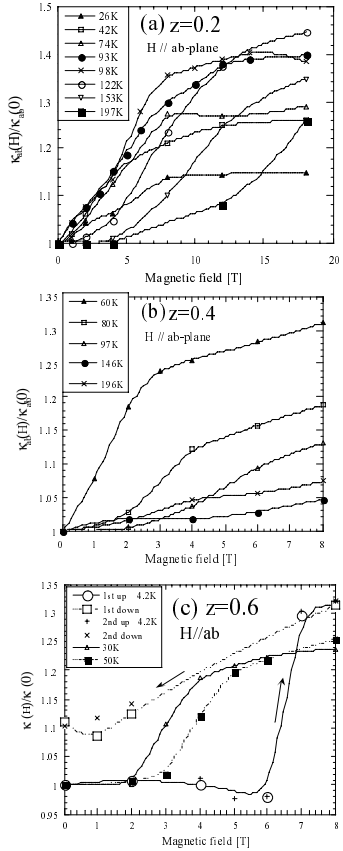}%
 \caption{\label{NMTC} normalized magnetothermal conductivity (NMTC) [$\kappa_{ab}$ (H)/$\kappa_{ab}$ (0)] of (La$_{1-z}$Pr$_{z}$) $_{1.2}$Sr$_{1.8}$Mn$_{2}$O$_{7}$ ($z=0$, 0.2, 0.4 and 0.6) single crystals.(a)$z$=0.2;(b) $z$=0.4 and (c) $z$=0.6. The field was applied in the $ab$-plane, parallel to the direction of the heat current. }
\end{figure}%

We discuss next, the normalized magnetothermal conductivity (NMTC) [$\kappa_{ab}$ (H)/$\kappa_{ab}$ (0)] of the Pr-substituted crystals at $z$=0.2 , $z$=0.4 and $z$=0.6 (Fig.\ref{NMTC}). The field was in the ab-plane and parallel to the direction of the heat current. Our data reveal several interesting features. First, a giant magnetothermal conductivity was observed in association with the occurrence of the CMR effect and the NMTC reached  about 30\%, up to 8T,  for all samples in spite of a lower $T_{C}$.  Here, it should be noted that  $\kappa_{c} (H)$/ $\kappa_{c}(0)$ is almost negligible in contrast with a rapid enhancement  of $\kappa_{ab}$ (H)/$\kappa_{ab}$ (0),  which is consistent with the absence of  anomalies observed in $\kappa_{c}(T)$ \cite{MA00}. 
Second, accompanying the field-induced first-order PI to FM transition of the insulating sample, z=0.6 , a clear hysteresis in the magnetothermal conductivity was observed.  The remnant NMTC at 4.2K reached nearly the same value of $\sim$10\% ,both in first and second scans, where, after the first scan, the sample was warmed up to high temperatures in order to demagnetize it. The value of NMTC of the $z=0.6$ sample  exhibited a step-like behavior at 4.2 K, while it showed a smooth rise with increasing temperature.  As previously pointed out, electronic contribution from the W-F law estimation using the CMR data reproduces neither an upturn in $\kappa_{ab}$  nor  a giant thermal effect.   Magnetostriction data around  $T_{C}$ showed a volume shrinkage upon application of a magnetic field, up to 8T, as well as spontaneous  striction ($\sim$0.1\%) upon  crossing  $T_{C}$ \cite{KI98,MA02}.  The negative magnetovolume effect  indicates a suppression in local lattice disorder;the reduction in phonon scattering due to structural disorder   leads to a giant magnetothermal effect  in the same way as the upturn  in $\kappa$ below  $T_{C}$. 
\begin{figure}
\includegraphics[width=10cm]{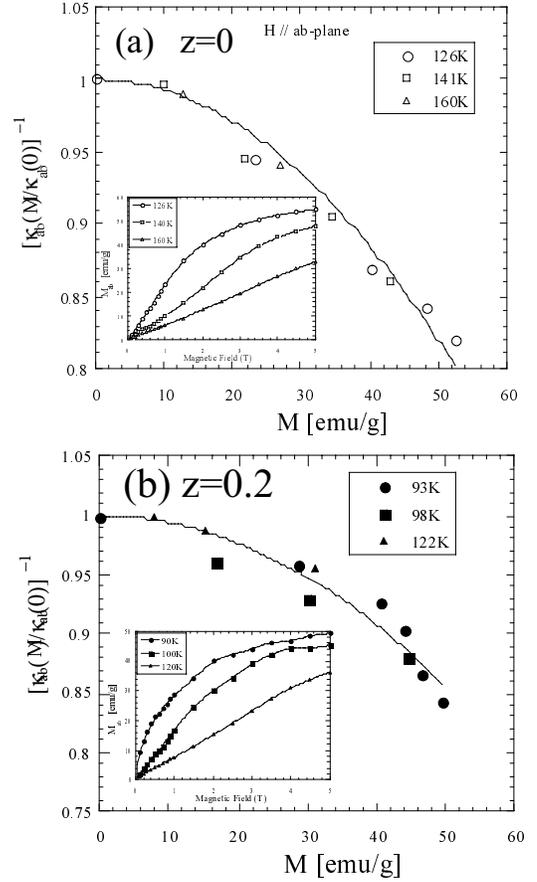}%
 \caption{\label{NMTR1}Normalized magnetothermal resistivity [$\kappa $$_{ab}(M)$/$\kappa$$_{ab}(0)$]$^{-1}$ of (La$_{1-z}$Pr$_{z}$)
 $_{1.2}$Sr$_{1.8}$Mn$_{2}$O$_{7}$ ($z=0$, 0.2, 0.4 and 0.6) single crystals as a function of magnetization, at selected temperature.(a)$z$=0;(b) $z$=0.2 and (c) $z$=0.4 and 0.6. The insets in Figs.5(a,b,c)represent $M_{ab}$-$H$ curves at the corresponding temperatures for the samples with $z$=0 , 0.2 and 0.6 respectively.(d)$z$ dependence of coefficient $A$ and lattice parameter $c/a$}
\end{figure}%

\begin{figure}
\includegraphics[width=10cm]{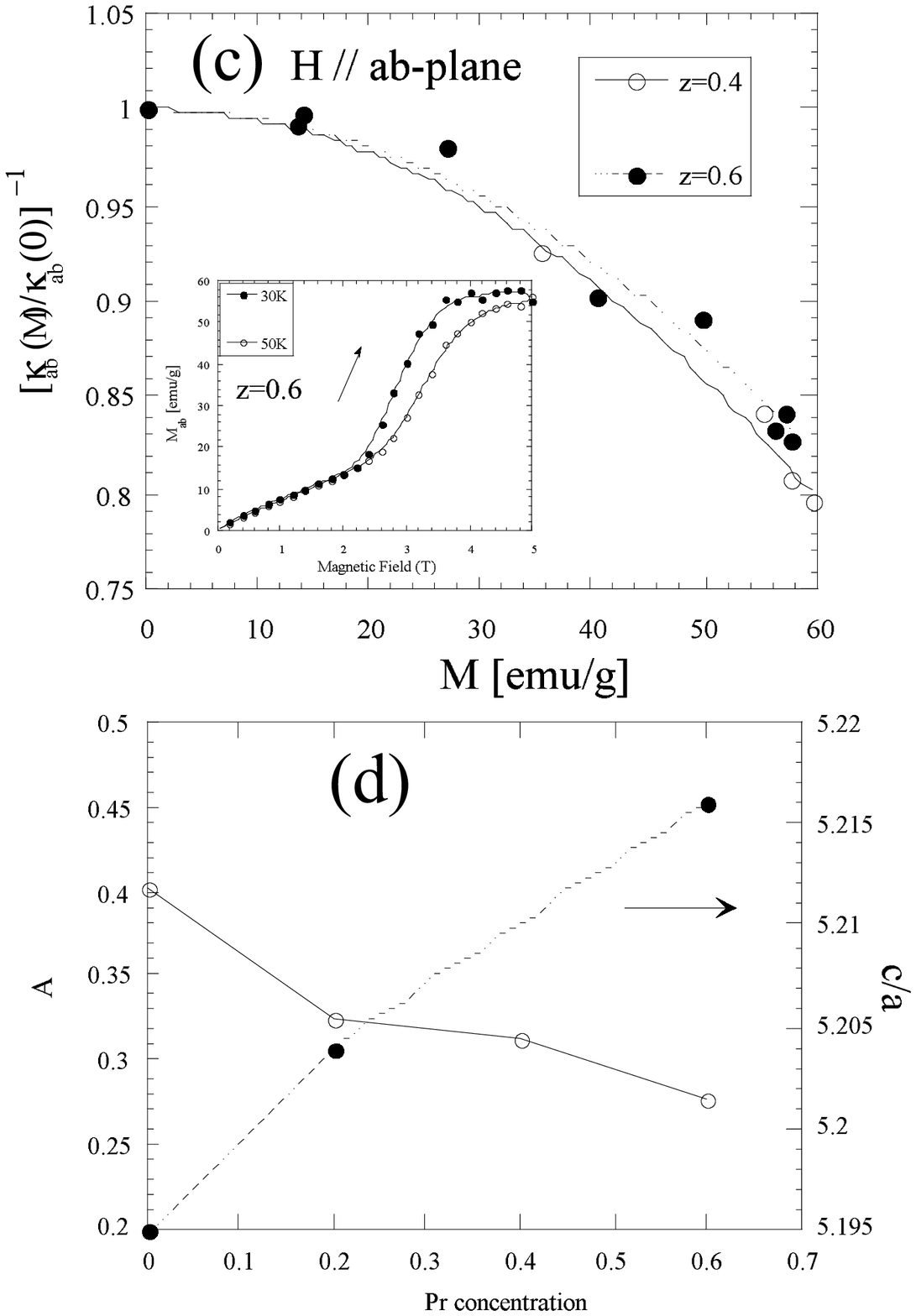}%
\label{NMTR2}
\end{figure}%

The normalized magnetothermal resistivity   [$\kappa $$_{ab}(M)$/$\kappa$$_{ab}(0)$]$^{-1}$ of our samples is plotted as a function of magnetization at selected temperatures in Fig.5 (a,b,c), 
where the insets in Figs.5 (a) , 5 (b) and 5 (c)represent $M_{ab}$-$H$ curves at the corresponding temperatures for the samples with $z$=0, 0.2 and 0.6 respectively.  It is easily found that the value of  the magnetothermal resistivity  is scaled with  a parabolic function of $1-A(M/M$$_{s}$)$^{2}$  ,as pointed out for cubic manganites ,   by  Cohn et al. \cite{CO97}. Here, $M_{s}$ is taken as the value of 3.6$\mu$$_{B}$ per Mn-sites.
The coefficient $A$ depends on the value of $z$ and shows a monotonical decrease from 0.4  for $z$=0 down to 0.28 for $z$=0.6, as shown in Fig5.(d).  This finding suggests a suppression  of the lattice-spin coupling in thermal CMR  due to Pr-substitution and can be explained as follows.   The Pr-substitution had a strong effect on the magnetic anisotropies. The low-temperature magnetic anisotropy $M_{ab}$/$M_{c}$ in a field of 100mT reduced  from $\sim$10 at $z$=0 to nearly 1 at $z$=0.2 and 0.4. For $z$=0.6, the easy axis of magnetization was along the $c$-axis. 
 On the other hand, the lattice parameters vary with Pr-substitution in such a manner that  the $c$-axis expands but the $a(b)$ axis shrinks (see Fig.5 (d)) which results in a change of $e_{g}$-electron character from planar $3d_{x^2-y^2}$ to linear $3d_{3z^2-r^2}$.  Accordingly,  such a variation of the $e_{g}$-electron character  accompanies that of magnetic anisotropy,  so that  Pr-substitution  probably suppresses indirectly the lattice-spin coupling in thermal CMR  through a variation of spin alignment, strongly coupled with orbital states. In other words, a higher magnetothermal conduction  favors  the occurrence of the easy axis of magnetization lying in the $ab$-plane.  It is an interesting feature that a change of $e_{g}$-electron character ,due to Pr-substitution, appears in  giant magnetothermal effect involving heat carrying phonons.

Finally, we shall stress  that at lower temperatures, far from $T_{C}$, a finite increase of $\kappa_{ab}$  ( $\sim$14\% at 8T)  for the $z$=0.2 crystal  was observed.  If we accept the above explanation, this finding indicates that phonon scattering due to local lattice disorder ( bond-length fluctuation)  remains, even in the metallic state, which is consistent with the anisotropic D-W data of La$_{2-2x}$Sr$_{1+2x}$Mn$_{2}$O$_{7}$ (0.32$\leq$x$\leq$0.4) in the metallic state.  Recently, Maezono et al. \cite{MAE02} pointed out that in manganite systems, the dynamic JT effect is more enhanced in the metallic region than in the insulating region  because of strong electron correlations.    Magnetothermal  measurements  ,at low temperatures , probe the lattice disorder not only  in the insulating region but also in the metallic region as the D-W factor  does.  Furthermore, it is desirable to clarify the crucial role played by the dynamic local lattice distortion on phonon conduction  ,not only in the insulating phase , but also in the metallic phase through magnetothermal measurements on high quality single crystal of the extremely doped bilayer compounds. 

In summary,  a systematic study of thermal conductivity as a function of temperature and magnetic field of single crystals of the compound (La$_{1-z}$Pr$_{z}$)$_{1.2}$Sr$_{1.8}$Mn$_{2}$O$_{7}$ for $z$(Pr) =0.2,0.4. and 0.6 was  performed. The $ab$-plane phonon conduction of the Pr-substituted $x=0.4$ samples is disturbed by the dynamical lattice distortions associated with polaron hopping on Mn sites.  On the other hands, the typical phonon behavior observed for the end member $x$=1 sample is probably attributed to  a reduction of the inhomogeneities in the local lattice distortions strongly coupled with small polarons.    The giant magnetothermal effect  has been explained  taking into account a systematic variation of the $e_{g}$-electron orbital states due to Pr-substitution.

%
\begin{acknowledgments}
The authors would like to thank Dr. H.Ogasawara for his technical support. This work was supported by a Grant-in-Aid for Scientific Research from the Ministry of Education, Science and Culture, Japan.
\end{acknowledgments}


\end{document}